\documentclass[doublespacing]{elsart}

\usepackage{graphicx}

\usepackage{amssymb}

\begin{document}

\begin{frontmatter}



\title{Thermodynamic and transport properties of underdoped cuprates from ARPES data}


\author[address1]{T. Yoshida\corauthref{cor1}}
\corauth[cor1]{Teppei Yoshida +1-650-725-5457(FAX)
\\ McCullough Bldg. Rm. 220, 476 Lomita Mall,
Stanford University, Stanford, CA 94305, USA}
\ead{teppei@stanford.edu}
\author[address1]{X. J. Zhou},
\author[address2]{H. Yagi},
\author[address1]{D. H. Lu},
\author[address2]{K. Tanaka},
\author[address2]{A. Fujimori},
\author[address3]{Z. Hussain},
\author[address1]{Z.-X. Shen},
\author[address4,address5]{T. Kakeshita},
\author[address6]{H. Eisaki},
\author[address5]{S. Uchida},
\author[address7]{Kouji Segawa},
\author[address7]{A.N. Lavrov} and
\author[address7]{Yoichi Ando}

\address[address1]{Department of Applied Physics and Stanford Synchrotron Radiation Laboratory, Stanford University, Stanford, CA94305, USA}
\address[address2]{Department of Physics and Department of Complexity Science and Engineering, University of Tokyo, Bunkyo-ku, Tokyo 113-0033, Japan}
\address[address3]{Advanced Light Source, Lawrence Berkeley National Lab, Berkeley, CA 94720, USA}
\address[address4]{Superconductivity Research Laboratory, ISTEC, Shinonome 1-10-13, Koto-ku, Tokyo 135-0062, Japan}
\address[address5]{Department of Physics, The University of Tokyo Tokyo 113-8656, Japan}
\address[address6]{National Institute of Advanced Industrial Science and Technology, Tsukuba 305-8568, Japan}
\address[address7]{Central Research Institute of Electric Power Industry, Komae, Tokyo 201-8511, Japan}

\begin{abstract}
The relationship between photoemission spectra of
high-$T_{\textrm{c}}$ cuprates and their thermodynamic and
transport properties are discussed. The doping dependence of the
expected quasi-particle density at the Fermi level
($E_\mathrm{F}$) are compared with the electronic specific heat
coefficient $\gamma$ and that of the spectral weight at
$E_\mathrm{F}$ with the in-plane and out-of-plane superfluid
density. We have estimated the electrical resistivity of
underdoped cuprates from the momentum distribution curve (MDC) at
$E_\mathrm{F}$ in the nodal direction. The temperature dependence
of the MDC width is also consistent with that of the electrical
resistivity.
\end{abstract}

\end{frontmatter}

\section{Introduction}

Thermodynamics and the transport properties of
high-$T_{\textrm{c}}$ cuprates are highly affected by the strong
electron correlation near the filling-control Mott transition. One
of the most unconventional thermodynamic properties is that the
electronic specific heat coefficient $\gamma$ shows pseudogap
phenomena in the underdoped region \cite{momono}. The electronic
resistivity show unconventional features such as $T$-liner
temperature dependence in the optimally doped region and metallic
behavior well exceeding the Ioffe-Regel limit in the lightly doped
region \cite{ando}. In order to elucidate  the origin of these
unconventional phenomena, angle-resolved photoemission
spectroscopy (ARPES) is an extremely powerful technique because
one can directly observe the spectral function $A({\bf
k},\omega)$, i.e., the electronic structure, near the Fermi level
($E_{\rm F}$). Therefore, information obtained from ARPES can be
directly compared with thermodynamic properties such as the
electronic specific heat. Also, the ARPES spectra near $E_{\rm F}$
give significant information about the transport properties. In
this article, we shall attempt to understand the unconventional
thermodynamics and transport properties based on ARPES spectra
mostly just at $E_{\rm F}$ \cite{yoshidaOD}.

\section{Thermodynamic properties}

The electronic specific heat coefficient $\gamma$ of a Fermi
liquid is given by $\gamma=\pi^2k_B^2/3 N(0)^*$ , where $N(0)^*$
is the quasi-particle (QP) density at $E_{\rm F}$. In a
single-band system like single-layer cuprates, each momentum
contributes one QP, if the system is a Fermi liquid. Therefore, if
one successfully fits the dispersion of the QP band to some energy
band model such as the tight-binding (TB) model, the density of
states (DOS) of that band yields $N(0)^*$.

In Fig. \ref{gamma}(a), the DOS determined by a TB fit to the
ARPES results of La$_{2-x}$Sr$_x$CuO$_4$ (LSCO) is compared with
the specific heat coefficient $\gamma$ of LSCO samples where the
$T_{\textrm{c}}$ is suppressed by Zn-doping \cite{momono}. The TB
parameters have been determined to reproduce the experimentally
observed Fermi surface and the slope of the band dispersions in
the nodal (0,0)-($\pi$,$\pi$) direction. Here, fine structures of
``kink" in the energy dispersions \cite{lanzara} are neglected in
the TB fit. The TB band DOS has a maximum around $x$=0.15 because
the flat band near ($\pi,0$) becomes closest to the Fermi level.
If the system is a Fermi liquid, the TB band DOS and $\gamma$
should be the same. However, the $\gamma$ decreases much faster
than TB band DOS with decreasing $x$. This difference indicates
that the QP density at $E_{\rm F}$ that would be expected for a
normal Fermi liquid is indeed depleted in the underdoped materials
and demonstrates how the opening of the pseudo-gap occurs around
($\pi,0$).

The doping dependence of the DOS of angle integrated photoemission
(AIPES) spectra at $E_\mathrm{F}$ are also shown in Fig.
\ref{gamma}(a). AIPES spectra have been obtained by integrating
the ARPES spectra with respect to $\mathbf{k}$ over the Brillouin
zone. As a whole, the AIPES DOS decreases with decreasing $x$ for
$x < 0.22$ due to the pseudo-gap opening. However, a closer
inspection reveals that the AIPES DOS decreases faster than
$\gamma$ with decreasing $x$. Note that the DOS at $E_\mathrm{F}$
measured by AIPES is given by $zN(0)^*$, where $z$ is the
renormalization factor. The faster decrease of the AIPES intensity
with decreasing $x$ is due to the effects of $z$, which also
decreases with decreasing $x$ \cite{Metallic}. From the above
comparison, one can say that the AIPES DOS reflects the effects of
both the pseudogap opening and the spectral weight renormalization
$z$.

Relationship between the ARPES spectral weight and the superfluid
density would also give important key about the superconductivity
\cite{feng}. The doping dependence of the ARPES spectral weight at
$E_\mathrm{F}$ is compared with that of the penetration depth
$1/\lambda^2\sim\rho_s/m^*$ \cite{panagopoulos} in Fig.
\ref{gamma}(b). The spectral weight at angle $\alpha$ (see inset)
is defined by the integrated spectral weight of the spectral
function $A({\bf k},\omega)$ at $E_\mathrm{F}$ over the momentum.

The decrease of both the in-plane and out-of plane superfluid
densities for $x<$0.20 is due to the pseudogap opening near
($\pi$,0). Particularly, the doping dependence of the out-of-plane
$(\rho_s/m^*)_c\sim1/\lambda_c^2$ well corresponds to that of the
spectral weight around $\sim$($\pi$,0) ($\alpha$=0$^\circ$). This
is because the out-of-plane $(\rho_s/m^*)_c\sim1/\lambda_c^2$ has
large matrix element around ($\pi$,0). On the other hand, the
nodal direction is expected to be the main contribution to
in-plane $(\rho_s/m^*)_{ab}\sim1/\lambda_{ab}^2$ because of the
lighter effective mass. However, the decrease with decreasing $x$
is somewhat different from the doping dependence of the nodal
spectral weight ($\alpha$=45$^\circ$). This implies that not only
the nodal QP's but also QP's over an extended $k$-region of the
``Fermi arc" condense into the superfluid.

\section{Transport properties}

The transport properties of high-$T_{\textrm{c}}$ cuprates show
unconventional behavior reflecting the strongly electron
correlation. Because of the pseudo-gap opening around ($\pi$,0),
the electronic states around the node mainly contribute to the
in-plane transport properties in the underdoped region. This
portion of the Fermi surface survives in the upderdoped compounds
as a ``Fermi arc" \cite{Metallic}. We shall estimate the in-plane
electrical resistivity from the observed ``Fermi arc" using the
conventional Drude formula $\rho=m^*/ne^2\tau$. Here, the
renormalized mass $m^*=\hbar k_\mathrm{F}/v_\mathrm{F}$ can be
determined from the quantities measured by ARPES, namely, the
Fermi velocity $v_\mathrm{F}$ and the  Fermi momentum
$k_\mathrm{F}$ of the hole-like Fermi surface centered at
($\pi,\pi$). From the MDC width $\Delta k$ at $E_\mathrm{F}$, the
mean free path is obtained as $l=1/\Delta k$, yielding the mean
free scattering time of $\tau=l/v_\mathrm{F}$. Thus, one can
evaluate $\rho=m^*/ne^2\tau=\hbar k_\mathrm{F}\Delta k/ne^2$ from
ARPES spectra.

Figure~\ref{Ektransport}(a) shows an ARPES intensity plot in the
$E$-$k$ space for  an LSCO ($x$ = 0.03) sample in the
(0,0)-($\pi,\pi$) direction. The energy dispersion determined by
the peaks in the MDC shows a ``kink" feature around 70 meV below
$E_\mathrm{F}$, which is presumably due to electron-phonon
coupling \cite{lanzara}. $v_\mathrm{F}\propto 1/m^*$ in the node
direction is almost doping independent $v_\mathrm{F}\sim 1.8$ eV
\textrm{\AA} for LSCO \cite{zhouNature}.

Now, we discuss how to estimate the carrier number $n$. Although
the Fermi-liquid picture yields the Fermi surface volume $\sim
1-x$, Hall coefficient results show unconventional behavior $n\sim
x$ in the underdoped region \cite{takagi}. The observed ``Fermi
arc" picture gives two possibilities to explain the evolution of
the carrier number like $n\sim x$ \cite{Metallic}. One possibility
is that the length of the ``Fermi arc" increases with $x$ and
yields $n\sim x$. However, the ``arc" length estimated from the
ARPES intensity does not change enough to explain the $n\sim x$
behavior. Another possibility is the evolution of the nodal
spectral weight with $x$. Indeed, the spectral weight as a
function of $x$ is remarkably similar to the $x$-dependence of the
Hall coefficient \cite{Metallic}.

Figure \ref{Ektransport}b shows the doping dependence of the
estimated in-plane electrical resistivity $\rho_\mathrm{ARPES}$
based on the Drude formula by using the parameters $k_\mathrm{F}$, $\Delta k$
and $n$ as described above. ARPES intensity $n_\mathrm{PES}\sim
n_\mathrm{Hall}$, $x$ (``small Fermi surface volume")
and $1-x$ (``large Fermi surface volume") are considered as the
carrier number $n$. $\rho_\mathrm{ARPES}$ evaluated using $n=x$ and
$n_\mathrm{PES}$ are in good agreement with the transport results
in the underdoped region. On the other hand, $\rho_\mathrm{ARPES}$
evaluated using $n=1-x$ is by more than one order of magnitude smaller
than the transport data in the underdoped region. This suggests
that the pseudogap and the small renormalized spectral weight is
the origin of the small carrier number and the high electrical
resistivity for the underdoped samples. On the other hand, the
discrepancy between $\rho_\mathrm{ARPES}$ ($n=x$,$n_\mathrm{PES}$)
and the transport data becomes pronounced in the optimally
doped region. This discrepancy may indicate a cross-over from the
Fermi arc to the large Fermi surface and/or may be due to the
contribution from the
 ($\pi$,0) region, which is not precisely taken into account in the
present simplified analysis. We leave such quantitative analysis
for future investigation. As a whole, the estimated $\rho_\mathrm{ARPES}$
for $n_\mathrm{PES}$ explains the doping dependence of
the transport data of the underdoped LSCO qualitatively well.

We also show the case of YBa$_2$Cu$_3$O$_y$(YBCO) in Fig.
\ref{YBCO}(a)(b). Relationship between the oxygen content $y$ and
the hole concentration $\delta$ has been estimated from the
electrical resistivity and the thermopower \cite{UchidaOng}. The
MDC peak width of YBCO is about twice as large as that of LSCO and
the MCD line shape of YBCO could not be well fitted to a single
Lorentzian. We consider that this is due to the bilayer splitting
in the nodal direction as predicted by band-structure calculations
\cite{Andersen}. We have therefore fitted the MDC to two
Lorentzians to estimate $\Delta k$, assuming that the chain band
does not exist at $E_\mathrm{F}$ for such low doping.
Fig.\ref{YBCO}(b) shows the estimated value of $\rho$ in the same
way as Fig.\ref{Ektransport}(b). While the transport data
\cite{ando_YBCO1,ando_YBCO2} is closer to $\rho_{\rm
ARPES}(n=\delta)$ for $\delta=0.04$, it approaches $\rho_{\rm
ARPES}(n=1-\delta)$ with increasing $\delta$, again implying the
crossover from the Fermi arc to the large Fermi surface.

Because the Fermi velocity $v_{\mathrm F}$ does not seem to change
with temperature, the temperature dependence of the inverse life
time $1/\tau$ is determined by that of $\Delta k=1/l=1/v_{\mathrm
F}\tau$. Figure \ref{dkT}(a) shows the temperature dependence of
MDC at $E_{\mathrm F}$ in the nodal direction for the $x$=0.07
sample. The MDC width $\Delta k$ is systematically broadened with
increasing temperature, indicating a decrease of the mean-free
path. As shown in Fig.\ref{dkT}(b), the temperature dependence of
$\Delta k$ well scales with that of the electric resistivity,
indicating that the temperature dependence of $\Delta k$
determines that of the resistivity. In optimally doped Bi2212, the
linear temperature dependence of the MDC width which scales with
resistivity was observed and discussed in the context of marginal
Fermi-liquid picture \cite{valla}. The present result also shows a
$T$-linear behavior in $\Delta k$ over a wide temperature range.
However, considering that electron-phonon interaction causes the
``kink" behavior \cite{lanzara} and contribute to the low energy
properties, the MDC width would be affected by phonons.

\section{Conclusion}
In summary, we have discussed the relationship between the ARPES
results for high-$T_{\textrm{c}}$ cuprates and their
thermodynamics and transport properties. From the comparison of
the ARPES and $\gamma$ and $\rho_s$, the pseudogap feature and the
spectral weight renormalization are discussed. We have estimated
the electric resistivity from the observed spectra and have
demonstrated that the obtained results explain the transport
properties. The temperature dependence of the MDC width $\Delta k$
in the nodal direction well scales with that of the electric
resistivity.

\begin{ack}
We are grateful to H. Fukuyama for enlightening discussions. This
work was supported by a Grant-in-Aid for Scientific Research
``Novel Quantum Phenomena in Transition Metal Oxides" from the
Ministry of Education, Science, Culture, Sports and Technology of
Japan and the New Energy and Industrial Technology Development
Organization (NEDO). Experimental data were recorded at ALS and
SSRL which are operated by the Department of Energy's Office of
Basic Energy Science, Division of Material Sciences. SSRL is also
operated by Division of Chemical Sciences.
\end{ack}

\clearpage \label{gamma}Fig. 1 (a) Comparison the electronic
specific heat coefficient $\gamma$ \cite{momono} and the TB band
DOS at $E_F$. The doping dependence of AIPES DOS at $E_\mathrm{F}$
is also shown. (b) Doping dependence of the spectral weight around
nodal and ($\pi$,0) regions. The in-plane and out-of-plane
superfluid densities deduced from the magnetic penetration depth
($\lambda_{ab}$, $\lambda_c$) \cite{panagopoulos} are compared
with them.

\label{Ektransport}Fig. 2 (a) Spectral intensity in the $E$-$k$
space along the nodal cut and the MDC at the Fermi level. (b)
Comparison of the electrical resistivity at 20 K \cite{ando} and
the resistivity estimated from ARPES. The carrier number $n$ is
assumed to be $n=x, n=1-x$ and $n=n_\mathrm{PES}$.

\label{YBCO}Fig. 3 (a) Spectral intensity of YBCO in the $E$-$k$
space along the nodal cut and the MDC at the Fermi level. (b)
Comparison of the electrical resistivity at 10 K
\cite{ando_YBCO1,ando_YBCO2} and the resistivity estimated from
ARPES. The carrier number $n$ is assumed to be $n=\delta$ and
$n=1-\delta$.

\label{dkT}Fig. 4 (a) Temperature dependence of the MDC at
$E_{\mathrm F}$ in the nodal direction for $x$=0.07. (b)
Comparison of the temperature dependence of the MDC momentum width
$\Delta k$ with that of the electrical resistivity.

\clearpage

\begin{figure}
\includegraphics[width=16cm]{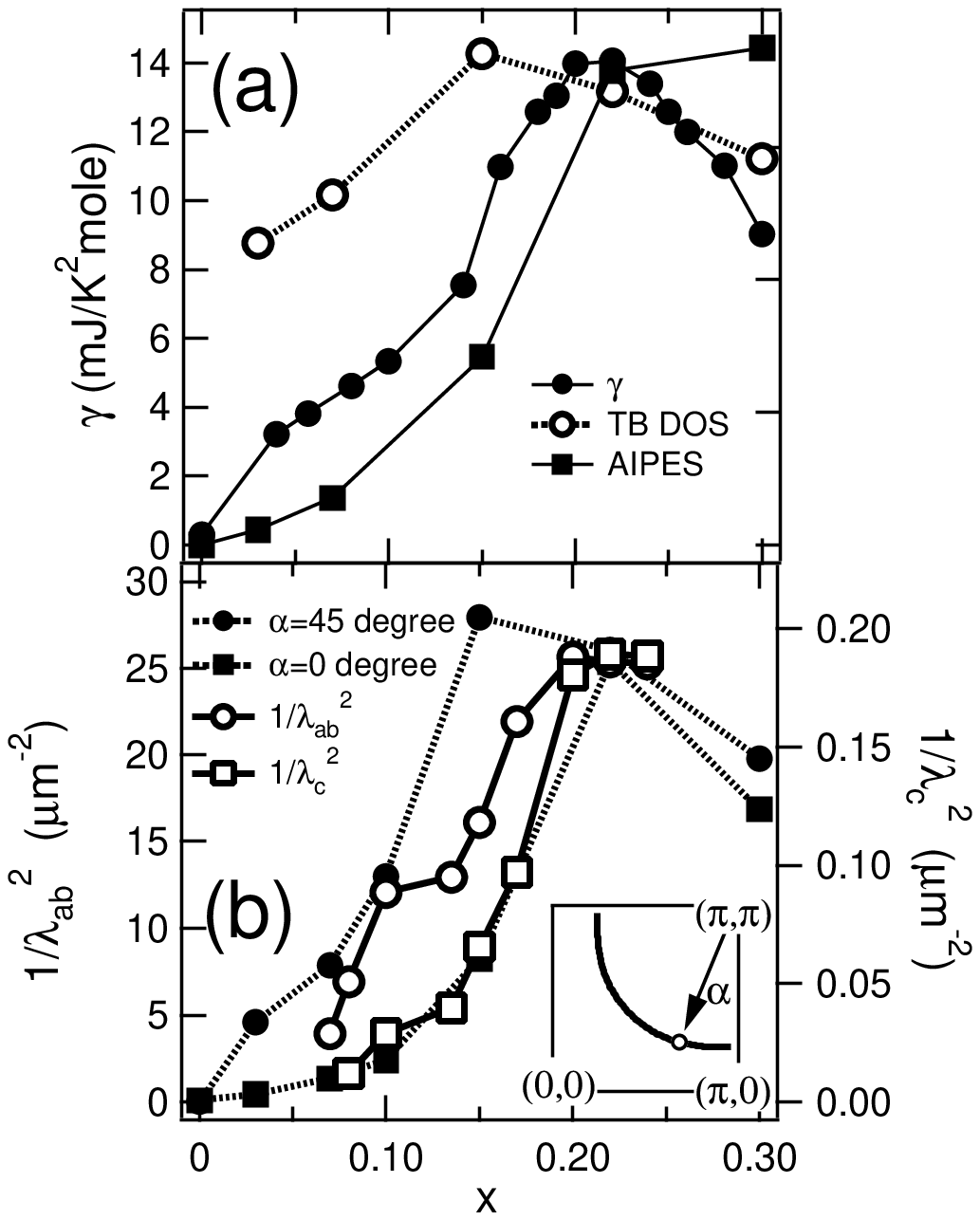}
\end{figure}
Fig. 1

\clearpage

\begin{figure}
\includegraphics[width=16cm]{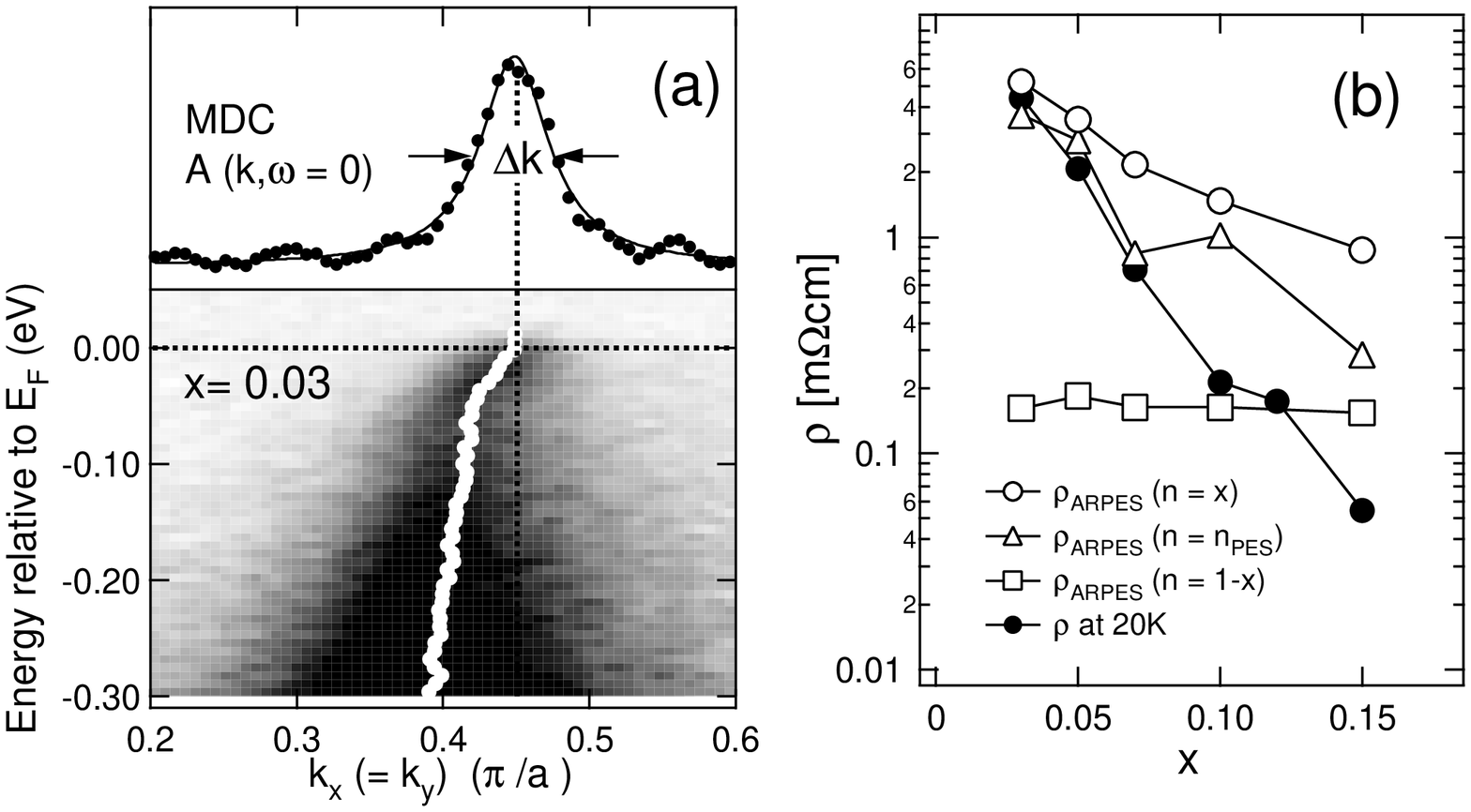}
\end{figure}
Fig. 2

\clearpage

\begin{figure}
\includegraphics[width=16cm]{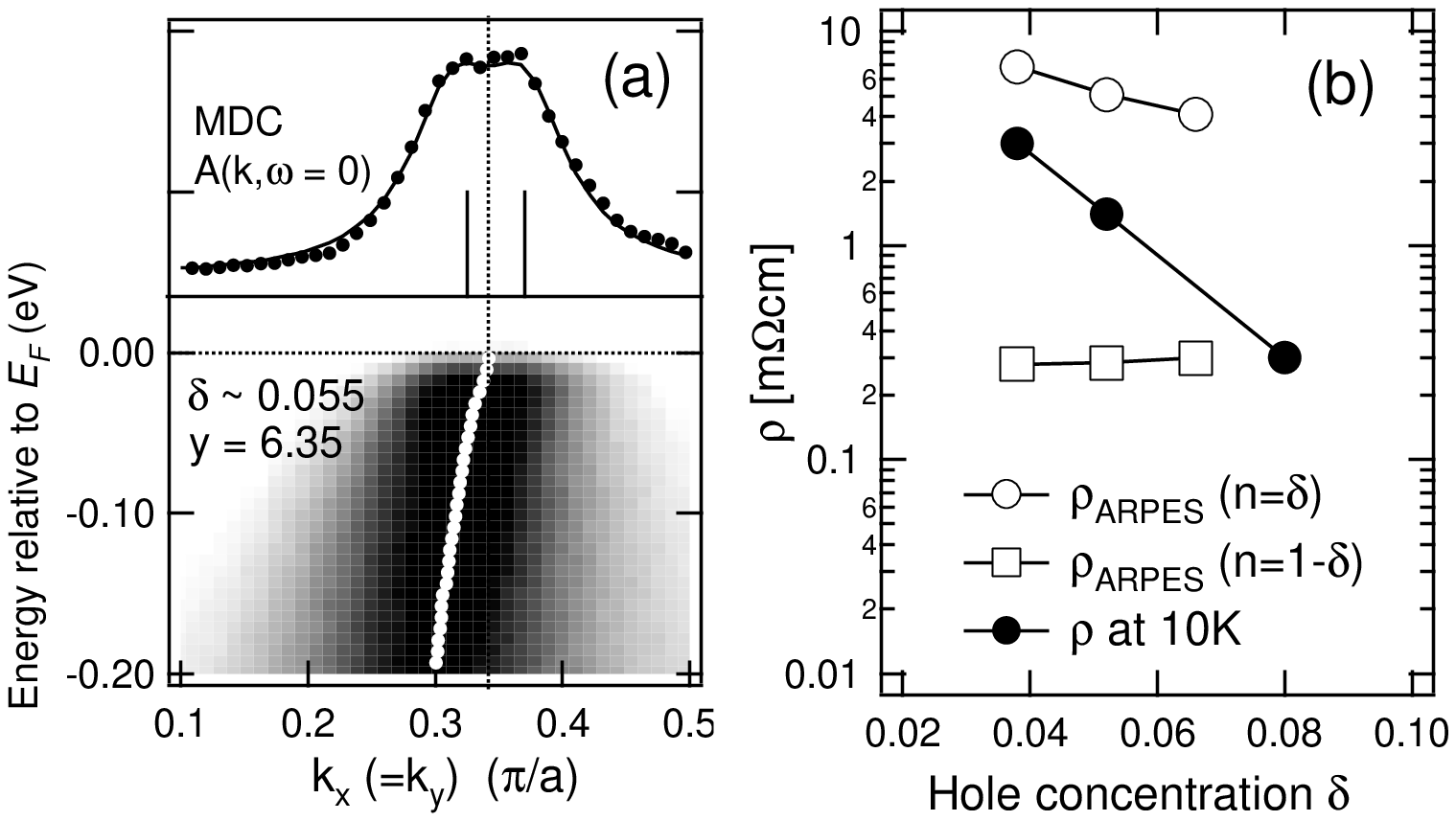}
\end{figure}
Fig. 3

\clearpage

\begin{figure}
\includegraphics[width=16cm]{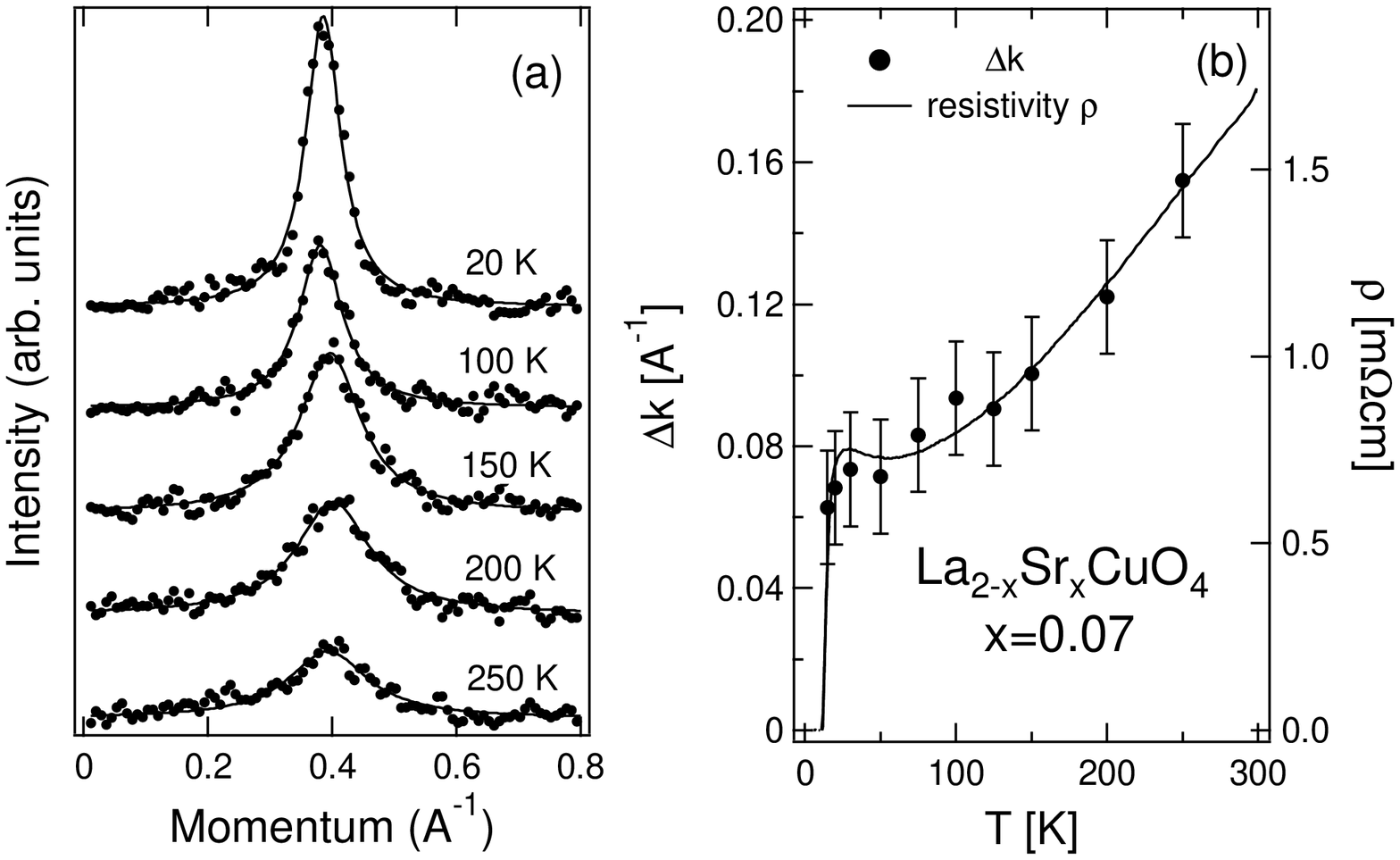}
\end{figure}
Fig. 4

\end{document}